\documentstyle[aps,prb]{revtex}
\begin{document}
\title{Spectral function of one hole in several one-dimensional spin
arrangements} 
\author{R.\ Hayn}
\address{Institute for Theoretical Physics, University of Technology, 
D-01062 Dresden, \\ 
and Institute for Solid State and Materials Research (IFW), 
D-01171 Dresden, Germany}
\author{R.\ O.\ Kuzian}
\address{Institute for Materials Science, Krjijanovskogo 3, 252180 Kiev,\\
Ukraine}
\date{\today}
\maketitle

\begin{abstract}
The spectral function of one hole in different magnetic states of the
one-dimensional $t$-$J$ model including three-site term and frustration
$J^{\prime}$ is studied. In the strong coupling limit $J \to 0$ 
(corresponding to $U \to \infty$ of the Hubbard-model) a set of eigenoperators
of 
the Liouvillian is found which allows to derive an exact expression for the
one-particle Green's function that is also applicable at finite temperature
and in an arbitrary magnetic state. 
The spinon dispersion of the pure $t$-$J$ model with the ground-state of the
Heisenberg model can be obtained by treating the corrections due to a small
exchange term by means of the projection method. The spectral function for the 
special frustration $J^{\prime}=J/2$ with the Majumdar-Ghosh wave function is
discussed in detail. Besides the projection method, a variational ansatz with
the set of eigenoperators of the $t$-term is used. We find a symmetric spinon
dispersion around the momentum $k=\pi/(2a)$ and a strong damping of the holon
branch. Below the 
continuum a bound state is obtained with finite spectral weight and a very
small separation from the continuum. Furthermore, the spectral function of the 
ideal paramagnetic case at a temperature $k_B T \gg J$ is discussed. 
\end{abstract}

\begin{pacs}
x71.27.+a, 73.20.Dx, 79.60.Bm
\end{pacs}

\thispagestyle{empty} \pagestyle{plain}

\section{Introduction}

The understanding of spin-charge separation is a central point in the
physics of low-dimensional electronic systems. That issue is most clearly
seen in one dimension where the exact solution of the 1D Hubbard model \cite
{LiWu} reveals that the low-energy physics is dominated by decoupled,
collective charge and spin excitations (also called holon and spinon, 
respectively). The idea that the spin and charge degrees of freedom separate
has also been proposed to explain the properties of 2D cuprate
superconductors. \cite{Anderson} 
However, even in one dimension, the Bethe ansatz solution did not yet give a
complete answer for the spectral function $A(k,\omega)$. Only for $U \to
\infty$ an exact expression for $A(k,\omega)$ was found \cite{SP} due to the
factorization property of the wave function according to Ogata and Shiba.
\cite{OgSh} The results for the insulating half-filled case \cite{SP} 
could also be generalized to other filling factors. \cite{penc}

The spin-charge separation was observed in ARPES measurements of the one
dimensional, dielectric cuprate SrCuO$_2$. \cite{K96} The spinon and holon
branch 
of the spectral function were seen, in contrast to the analogous experiment of
one hole in the CuO$_2$ plane \cite{wells} where spin and charge are coupled
and the spin polaron quasiparticle has a dispersion proportional to $J$ as 
proposed theoretically in Refs.\ \onlinecite{KLR,MH,EB}. The ARPES
spectra in SrCuO$_2$ were analyzed using the pure $t$-$J$ model. On the
other hand, many 1D compounds, like for instance CuGeO$_3$, \cite{F98} are
characterized by frustration in the magnetic subsystem which may lead to a
gap in the spin excitation spectrum. For the special frustration $ 
J^{\prime}/J=0.5$ and in the limit $J \to 0$ an analytic expression for $ 
A(k,\omega)$ was derived recently, \cite{Maekawa} under the assumption that
the wave function factorizes. Besides the frustration, also temperature
effects are important as it was observed in ARPES measurements on 
Na$_{0.96}$V$_2$O$_5$. \cite{K99} So, there is a clear need to study the
spectral 
function systematically under the influence of frustration and temperature
and to derive analytic expressions.

The present work focuses mainly on the effect of frustration and temperature
on the spectral function in the insulating case. For that, we rederive first
the exact solution of Sorella and Parola \cite{SP} in a straightforward
way using Green's function technique. That is possible due to our finding of
a set of eigenoperators of the Liouvillian (Sec.\ III) in the strong coupling
limit $J \to 0$. As a consequence, our
derivation is applicable for any magnetic state and any temperature in that
limit. Especially, one can show that the result which was
derived in Ref.\ \onlinecite{Maekawa} does not depend on the assumption that
the wave function factorizes. We present analytic expressions for the spectral
function of one hole in several magnetic 
states: (i) the ground-state of the antiferromagnetic
Heisenberg model, (ii) the Majumdar-Ghosh wave function \cite{MG} at the
special 
frustration $J^{\prime}/J=0.5$, and
(iii) the ideal paramagnetic state at temperatures much larger than the
exchange energy $k_B T \gg J$. 

For large finite coupling we compare two methods to account for
corrections $J \propto (t^2/U)$, namely the projection method and a
variational ansatz using the set of eigenoperators of the $t$-term. We 
show that the former method yields a reasonable description of the spinon
dispersion in the pure $t$-$J$ model (Sec.\ IV) and an approximate result
for the spectral function of the Majumdar-Ghosh model. For $J^{\prime}=J/2$ it 
misses the bound state below the continuum which is obtained by the more
accurate variational method (Sec.\ VB). The bound state has a finite spectral 
weight but a very small separation from the continuum. Both methods show in
the Majumdar-Ghosh case that the low energy region for momenta $k$ between
$\pi/2$ and $\pi$ (lattice constant $a=1$) will be filled with states, that
the spinon dispersion (i.e.\ that 
collective excitation corresponding to the lower edge of the continuum)
becomes symmetric around $\pi/2$, and they indicate an
overdamped holon branch. The damping of the holon 
branch is extremely large for very high temperatures (Sec.\ VI). 

Before presenting our results let us shortly discuss the different
understandings of the term ``spin-charge separation'' as it can be met in
the literature. The naive picture means that the low energy effective
Hamiltonian may be written as
\begin{equation}
\label{hhs}\hat{H}=\hat{H}_h+\hat{H}_s \; , \quad  
[\hat{H}_h,\hat{H}_s]=0 \; , 
\end{equation}
and the electron operator is the product
\begin{equation}
\label{sh}c_{i\sigma }=s_{i\sigma }h_i^{\dagger } \; ,
\end{equation}
where spinon $s$ and holon $h$ can be basically regarded as free particles.
Then the normalized (i.e. $\int A(k,\omega )d\omega =1$) spectral function
is
\begin{equation}
\label{Astr}A(k,\omega )=\frac 1L\sum_Q2f(Q)\delta \left[ \omega
-\epsilon_h(k-Q)-\epsilon_s(Q)\right] ,
\end{equation}
where $f(Q)=\theta (\frac \pi 2 - \left| Q \right|)$ is the Fermi distribution
function of spinons, $\theta (x)$ is the Heaviside step function, $L$ is the
number of sites, $\epsilon_h,\epsilon_s$ being holon and spinon energies,
respectively.

However, the naive understanding is not that one which is realized in 1D
electron systems. \cite{Nagaosa} There, it was found that the eigenstates
factorize in the limit $U \to \infty$ in the form
\begin{equation}
\label{wf}\psi (x_1,\ldots ,x_N,y_1\ldots y_M)=\psi _{SF}(x_1,\ldots
,x_N)\phi _H(y_1\ldots y_M),
\end{equation}
where $x_1,\ldots ,x_N$ are the spatial coordinates of the $N$ electrons on
a $L$-site ring, and the $y_1\ldots y_M$ 'coordinates' label the position of
the spin-up electrons on the squeezed Heisenberg ring, i.e.\ on the $N$
occupied sites. The $\psi _{SF}$ is a spinless fermion state, and $\phi _H$
is an eigenstate of an $N$-site Heisenberg Hamiltonian with periodic
boundary conditions. The product form of equation (\ref{wf}) should not be
interpreted as a trivial decoupling between charge and spin. In fact, the
momentum of the spin wave function imposes a twisted boundary condition on
the spinless fermion wave function. As a result, the Fermi distribution
function $f(Q)$ in (\ref{Astr}) will be replaced by a function $Z(Q)$ that
is the expectation value of a chain of spin operators that has to be
determined from the pure spin system \cite{SP} (for the details see Sec.\
III). The singularity of $Z(Q)$ produces additional peaks in $A(k,\omega)$.
This correct answer for the spectral function may be understood as a
manifestation of the {\em phase string} effect. \cite{Weng} It means that
spinon and holon interact with each other via a nonlocal phase-string.
Instead of (\ref{sh}) we should write
$$
c_{i\sigma }=s_{i\sigma }h_i^{\dagger }\exp \left[ \frac \pi 2 
\sum_{l>i}h_l^{\dagger }h_l+\frac \pi 2\sum_{l>i}(s_{l\sigma }^{\dagger
}s_{l\sigma }-1)\right] .
$$
It should be noted that the phenomenon of spin-charge separation is not
restricted to the limit $U \to \infty$ in the 1D Hubbard model. At any
finite $U$ the spin and charge fluctuations propagate with different
velocities. \cite{LiWu} That means that after some time the spin and charge
degrees of freedom will be separated in space. But there is no analytic
solution for the spectral function of the Hubbard model at arbitrary values of
$U$ and also the present calculation treats terms of order $(t^2/U)$ as a
perturbation. In that sense we will understand here spin-charge separation as
a manifestation of the factorization property (\ref{wf}) in the spectral
density. Sharp maxima in the continuum correspond to collective excitations
whereas possible bound states indicate special eigenfunctions with a strong
coupling between spin and charge. Another possible effect of additional terms
in the Hamiltonian is the broadening (i.e.\ the damping) of the collective
excitations. 

\section{Model and spectral density}

To describe the low energy physics of compounds with a 1D electronic
structure it is sufficient in most cases to take into account only that 
band which 
is closest to the Fermi energy (see for instance Ref.\ \onlinecite{RoDre}).
Treating the on-site Coulomb interaction explicitly, one obtains the well
known 1D Hubbard model. In the present calculation we restrict
ourselves to the strong coupling limit $U\gg t$ where we may project out the
subspace of doubly occupied sites, and for the lower Hubbard band we obtain
the effective Hamiltonian
\begin{equation}
\label{Heff}\hat H=\hat t+\hat J+\hat t_3 \; ,
\end{equation}
where
\begin{equation}
\label{tx}\hat t=-t\sum_{i,g,\alpha}X_i^{\alpha 0}X_{i+g}^{0\alpha } \; ,
\end{equation}
\begin{equation}
\label{Jx}\hat J=\frac J2\sum_{i,\alpha,\beta}X_i^{\alpha \beta
}X_{i+1}^{\beta \alpha } \; ,
\end{equation}
\begin{equation}
\label{t3x}\hat t_3=t_3\sum_{i,g,\alpha,\beta}X_i^{\alpha 0}X_{i+g}^{\beta
\alpha}X_{i+2g}^{0\beta } \; ,
\end{equation}
and $\alpha ,\beta =\uparrow \downarrow $; $g$ are the nearest neighbors $ 
g=\pm 1$. The Hamiltonian is valid near half filling
($X_i^{++}+X_i^{--}=1$). The parameters $J$ and $t_3$ are connected with the
original values of 
the Hubbard model by
\begin{equation}
\label{tJrel}J=4t_3=4t^2/U \, ,
\end{equation}
but the Hamiltonian (\ref{Heff}) is more general, if we relax the condition 
(\ref{tJrel}). It may be derived directly from the more realistic three-band
Hubbard model. Then, the $t_3$-term often becomes negligible and one
obtains the $t$-$J$ model. The Hamiltonian (\ref{Heff}) is written in terms
of Hubbard projection operators that act in the subspace of on-site states
\begin{equation}
\label{xab}X_i^{\alpha \beta }\equiv \left| \alpha ,i\right\rangle
\left\langle \beta ,i\right| \; , \quad 
\alpha ,\beta =0,\uparrow ,\downarrow ,2 \; .
\end{equation}
They are related with bare fermionic and spin operators through
\begin{equation}
\label{xs2}X_i^{\sigma 0}= c_{i,\sigma}^{\dagger} (1-n_{i,-\sigma }) \quad ,
\quad X_i^{\sigma 2}=-\sigma c_{i,-\sigma }n_{i,\sigma } \; ,
\end{equation}
\begin{equation}
\label{xss}X_i^{+-}=S_i^{+}=c_{i,\uparrow }^{\dagger }c_{i,\downarrow }
\quad , \quad X_i^{\sigma \sigma }=\frac 12+\frac \sigma 2\left(
c_{i,\uparrow }^{\dagger }c_{i,\uparrow }-c_{i,\downarrow }^{\dagger
}c_{i,\downarrow }\right) =\frac 12+\sigma S_i^z \; ,
\end{equation}
with $\sigma=\pm 1$. Other relations are easy to obtain with the use of the
main property of 
Hubbard operator algebra
\begin{equation}
\label{xalg}X_i^{\alpha \beta }X_i^{\gamma \lambda }=\delta _{\beta \gamma
}X_i^{\alpha \lambda } \; ,
\end{equation}
that follows immediately from the definition (\ref{xab}). The commutation
relations for operators on different sites are fermionic for operators that
change the number of particles by odd integers, like  (\ref
{xs2}), and bosonic for others. In the presence of frustration in the
magnetic system, which is discussed for instance for CuGeO$_3$, the $t$-$J$
Hamiltonian may be generalized by inclusion of the $J^{\prime }$-term
\begin{equation}
\label{Jprime}\hat J^{\prime}=\frac{J^{\prime}}{2} \sum_{i,\alpha, 
\beta}X_i^{\alpha \beta }X_{i+2}^{\beta \alpha } \; .
\end{equation}

Our aim is to calculate the one-particle two-time retarded Green's function $ 
G(k,\omega )$ and the spectral density of one hole in the magnetic state 
\begin{equation}
\label{A}A(k,\omega )=-\frac 1\pi {\rm Im}G(k,\omega + i 0^{+}),
\end{equation}
that is roughly proportional to the ARPES signal intensity. We define
\begin{equation}
\label{rGf}2\pi \delta (k-k^{\prime })G(k,\omega )=\langle\langle
X_k^{\sigma 0}|X_{k^{\prime }}^{0\sigma} \rangle\rangle \equiv -i
\int_{t^{\prime }}^\infty \!\!dte^{i \omega (t-t^{\prime })}\langle
\{X_k^{\sigma 0}(t),X_{k^{\prime}}^{0\sigma} (t^{\prime })\}\rangle \; ,
\end{equation}
where
$$
X_k^{\sigma 0}=\sqrt{2}\sum_{m=-\infty }^{+\infty }{\rm e}^{-i
km}X_m^{\sigma 0} \; ,
$$
$$
\left\langle \left\{ X_k^{\sigma 0},X_{k^{\prime }}^{0\sigma}\right\}
\right\rangle =2\pi \delta (k-k^{\prime }) \; ,
$$
and where $\{ \dots , \ldots \}$ means the anticommutator. The expectation
value denotes the thermal average over a grand canonical ensemble:
\begin{equation}
\label{avdef}\langle ...\rangle =Q^{-1}\mbox{Sp} \ [{\rm e}^{-\beta (\hat H 
-\mu {\hat N})}...],\ Q= \mbox{Sp} \ {\rm e}^{-\beta (\hat H-\mu {\hat N})}.
\end{equation}
Here Sp implies taking the trace of an operator, ${\hat N}$ is the particle
number operator, $\beta =(kT)^{-1}$ is an inverse temperature, and $\mu $
represents the chemical potential. The time dependence of the operator $B(t)$
is given by 
$B(t)={\rm e}^{it({\hat H}-\mu {\hat N})}
B{\rm e}^{-it({\hat H}-\mu {\hat N})}$.

\section{Eigenoperator and holon dispersion}

Let us consider first the limit $U\rightarrow \infty $ in the Hubbard model
or $J,t_3\rightarrow 0$ in (\ref{Heff}). Then only the $t$-term
$$
\hat t=-t\sum_{i,g,\alpha}X_i^{\alpha 0}X_{i+g}^{0\alpha } \; ,
$$
is nonzero. Note that it is a true many-body Hamiltonian due to the
constraint of no double occupancy, as we see from Eq.\ (\ref{xs2}). We
introduce the set of operators
\begin{equation}
\label{vmr}v_{m,r}= \sum_{\alpha_1,\ldots,\alpha_r} X_m^{\sigma \alpha
_1}X_{m+g}^{\alpha _1\alpha _2}\ldots X_{m+r-g}^{\alpha _{r-1}\alpha
_r}X_{m+r}^{\alpha _r0} \; , \qquad g={\rm sign} \ (r) \; ,
\end{equation}
for which
\begin{equation}
\label{emvmr}\left[ v_{m,r},\hat t\right] =t\left(
v_{m,r-1}+v_{m,r+1}\right)
\end{equation}
holds at half filling. Any operator (\ref{vmr}) can be considered as a string
operator of a 
certain length consisting of a hole and an attached string of spin flips. Such 
a string can be produced by creating a hole in the N\'{e}el state and applying
several times the kinetic energy (\ref{tx}) which creates misaligned
spins. Similar string operators were used to describe the spin polaron
quasiparticle in the 2D case. \cite{EB,HRY}
We make double Fourier transform
\begin{equation}
\label{eigop}v_{k,q}=\sqrt{2}\sum_{m,r=-\infty }^{+\infty }{\rm e}^{-i
km-i qr}v_{m,r} \; , 
\end{equation}
and see that
\begin{equation}
\label{hdisp}\left[ v_{k,q},\hat t\right] =2 t v_{k,q} \cos q \; .
\end{equation}
The interpretation in terms of the string operator (\ref{vmr}) is quite
easy. We see that only the right end of the operator (\ref{vmr}) was
influenced by the $t$-term. Therefore, we may identify the right end of
$v_{m,r}$ with the holon excitation. In the next Section, it will become clear 
that the left end of $v_{m,r}$ may be connected with the spinon. 

The operators $v_{k,q}$ (\ref{eigop}) are eigenoperators of the
Liouvillian $\cal{L}$ of the problem, where  
$
{\cal L} \hat{A} \equiv [ \hat{H}, \hat{A} ] 
$. 
Note that it is one of the rarest, if not the
unique case in many-body physics that the explicit form for a set of 
eigenoperators can be given. From (\ref{hdisp}) we see that the equation of
motion for the corresponding string operator Green's function closes and it
has a simple pole form
\begin{equation}
\label{hgf}\langle\langle v_{k,q}|v_{k^{\prime },q^{\prime }}^{\dagger }
\rangle\rangle =\frac{\left\langle \left\{ v_{k,q},v_{k^{\prime },q^{\prime
}}^{\dagger } \right\} \right\rangle }{\omega -2t\cos q} \; ,
\quad \quad
\left\langle \left\{ v_{k,q},v_{k^{\prime },q^{\prime }}^{\dagger }\right\}
\right\rangle =8\pi ^2\delta (k-k^{\prime })\delta (q-q^{\prime })Z(k-q+\pi)
\; ,
\end{equation}
where the spectral weight is the expectation value
\begin{equation}
\label{Kq}
Z(q+\pi)=\frac{1}{2}\sum_{r=-\infty }^{+\infty }{\rm e}^{-i qr}\left\langle
\Omega_{r}\right\rangle \; ,
\end{equation}
of a chain of $X$-operators
\begin{eqnarray}
\Omega _{r} &=&
\sum_{\alpha_1,\ldots,\alpha_r,\sigma}
X_m^{\sigma \alpha_1}X_{m+g}^{\alpha_1 \alpha_2}
\ldots X_{m+r-g}^{\alpha_{r-1} \alpha_r}X_{m+r}^{\alpha_r \sigma}
\nonumber \\
&=&(2{\bf S}_m{\bf S}_{m+g}+\frac 12)(2{\bf S}_{m+g}{\bf S}_{m+2g}+\frac
12)\ldots (2{\bf S}_{m+r-g}{\bf S}_{m+r}+\frac 12) \; ,
\label{25s}
\end{eqnarray}
to be calculated for the pure spin-system without any hole.
The expectation value $\langle \Omega_r \rangle$ of (\ref{25s}) 
cannot depend on the starting point $m$ due
to the translational symmetry of the problem. 
The operator (\ref{25s}) was introduced
in Ref.~\onlinecite{OgSh} and explicit values on a 26-site Heisenberg ring
were given for $T=0$, when $\langle \ldots \rangle$ becomes the average over
the ground-state. Asymptotically, the following behavior was found
\begin{equation}
\label{e25}
\langle \Omega_{l} \rangle \to \frac{1}{\sqrt{l}} \mbox{Re} \left[ A
{\rm e}^{i \pi l/2} \right]
\end{equation}
which leads to a square root singularity of $Z(Q)$. Using additionally the
exact values $\langle \Omega_{0} \rangle = 1$ and $\langle \Omega_{1} \rangle
= 1 - 2 \ln 2$ the following formula may be derived \cite{SP} 
\begin{equation}
\label{ZHeis}Z(Q) = \left( -0.393+0.835/\sqrt{\cos Q}\right) \theta 
(\frac \pi 2 - \left| Q \right| ) \; . 
\end{equation}
For the hole Green's function (\ref{rGf})  we have
\begin{equation}
\label{xkgf}
\langle\langle X_k^{\sigma 0}|X_{k^{\prime }}^{0\sigma}\rangle\rangle
=\int_{-\pi }^{+\pi }\frac{dq}{2\pi }\int_{-\pi }^{+\pi }\frac{dq^{\prime }}{ 
2\pi } \langle\langle v_{k,q}|v_{k^{\prime },q^{\prime }}^{\dagger
}\rangle\rangle
= 2\pi \delta (k-k^{\prime })\int_{-\pi }^{+\pi }\frac{dq}{\pi } 
\frac{Z(k-q+\pi)}{\omega -2t\cos q},
\end{equation}
and the spectral density is obtained in the way
\begin{equation}
\label{AJ0}
A(k,\omega )=\int_{-\pi }^{+\pi }\frac{dQ}{\pi }Z(Q)\delta
\left[ \omega +2t\cos (k-Q)\right] \; .
\end{equation}
That gives the exact answer in the strong coupling limit $(J \to 0)$ where
only the holon dispersion $\epsilon_h(q)=-2t \cos q$ is important. But the
ratio $J^{\prime}/J$ may be arbitrary and (\ref{AJ0}) is not only exact in
the Heisenberg case with $Z(Q)$ from (\ref{ZHeis}). Instead, from our
derivation follows its validity for arbitrary magnetic states and it is not
restricted to zero temperature. Then, however, $Z(Q)$ is different. In the
following we will give exact results for i) the Majumdar-Ghosh wave function
at the special frustration $J^{\prime}/J=0.5$, and ii) the ideal paramagnetic
case at $k_B T \gg J$. Two cases are quite trivial, namely the saturated
ferromagnetic 
case and the classical N\'{e}el state. The former one leads to
\mbox{$Z(Q+\pi)\propto\delta(Q)$} and a spectral function like for free
fermions,  
whereas the latter case leads to the Brinkmann-Rice continuum \cite{BR}
($Z(Q)=1/2$). A magnetic state inbetween the classical N\'{e}el and
the Heisenberg case could, in principle, also be considered, for which the
one-hole spectral 
function was derived in Ref.\ \onlinecite{Brenig} treating the
spin-fluctuations as perturbation. 

\section{Spinon dispersion}

In real systems the ratio $J/t$ is roughly 0.3. It means that they are in
the regime of strong coupling and the above consideration correctly describe
the largest energy scale $\propto t$. Now, we want to estimate the
corrections that arise from other terms of the Hamiltonian (\ref{Heff}).
First we note that $v_{k,q}$ are eigenoperators for the $t_3$-term
\begin{equation}
\label{t3disp}\left[ v_{k,q},\hat t_3\right] =-2 t_3 v_{k,q} \cos 2q \; ,
\end{equation}
which leads to the replacement $\epsilon_h(q) \to \epsilon_h(q) + 2 t_3 \cos
2q$ in the denominator of (\ref{hgf}). The commutation with $\hat J$ gives
(see Appendix A for the details)
\begin{equation}
\label{comJ}\left[ v_{k,q},\hat J\right] = - J \cos (k-q) v_{k,q}+\frac J2 
(v_{k,q}^{\prime }+v_{k,q}^{\prime \prime }) \; ,
\end{equation}
where $v_{k,q}^{\prime }$ and $v_{k,q}^{\prime \prime }$ are Fourier
transforms of the operators
\begin{equation}
\label{nupr}v_{m,r}^{\prime } = \sum_{\gamma,\alpha_1,\ldots,\alpha_r}
X_m^{\sigma \gamma}X_{m-g}^{\gamma \alpha_1}X_{m+g}^{\alpha _1\alpha _2}
\ldots X_{m+r-1}^{\alpha _{r-1}\alpha _r}X_{m+r}^{\alpha _r0} \; ,
\end{equation}
\begin{equation}
\label{nu2pr}v_{m,r}^{\prime \prime
}=-\sum_{\gamma,\alpha_1,\ldots,\alpha_r} X_m^{\sigma \alpha
_1}X_{m+g}^{\alpha _1\alpha _2}\ldots X_{m+r-g}^{\alpha _{r-1}\alpha
_r}X_{m+r+g}^{\alpha _r\gamma }X_{m+r}^{\gamma 0} \; .
\end{equation}
It is impressive that terms, which come from the commutation of ``inner''
$X_n$ operators in $v_{m,r}$ with $n$ between the points $m$ and $m+r$ cancel 
each other and only the terms coming from the ends remain. The term
$v_{m,r}^{\prime\prime}$ presents a distortion of the right end of $v_{m,r}$
by means of the exchange part and may be interpreted as the loss of magnetic
energy due to the presence of a holon. On the other hand, the term
$v_{m,r}^{\prime}$, with a distorted left end will be shown to give rise to
the spinon dispersion. We really observe the ``separate'' motion of the holon
that is represented by the right end of $v_{m,r}$ and of the spinon that is
the left end of $v_{m,r}$. The holon motion is governed by the $t$-term
and the spinon motion by the  
$J$-term. We put the word ``separate'' in quotes because the motion 
remains correlated due to the set of ``inner'' $X_n$ operators, connecting the
ends of $v_{m,r}$.

We need an approximate approach to account for
$v_{k,q}^{\prime}+v_{k,q}^{\prime \prime }$. For this purpose we use the
projection technique
\begin{equation}
\label{ptdec}
v_{k,q}^{\prime }+v_{k,q}^{\prime \prime }\approx \frac{ 
\left\langle \left\{ v_{k,q}^{\prime }+v_{k,q}^{\prime \prime
},v_{k,q}^{\dagger }\right\} \right\rangle }{\left\langle \left\{
v_{k,q},v_{k,q}^{\dagger }\right\} \right\rangle }v_{k,q}\;.
\end{equation}
Now, the Green's function for the string operator has the form
\begin{equation}
\label{nugf}\langle \langle v_{k,q}|v_{k^{\prime },q^{\prime }}^{\dagger
}\rangle \rangle =\frac{8\pi ^2\delta (k-k^{\prime })\delta (q-q^{\prime
})Z(k-q+\pi)}{\omega -2t\cos q+2t_3\cos 2q-\epsilon _s(k-q)},
\end{equation}
where $\epsilon _s(k-q)$ is defined by the equation
\begin{equation}
\label{sdisp}
\left\langle \left\{ \left[ v_{k,q},\hat J\right] ,v_{k^{\prime
},q^{\prime }}^{\dagger }\right\} \right\rangle \equiv 8\pi ^2\delta
(k-k^{\prime })\delta (q-q^{\prime })\epsilon _s(k-q)Z(k-q+\pi) \; .
\end{equation}
The contribution of the $v_{m,r}^{\prime\prime}$ term to the spinon dispersion 
is determined by an expression of the form
\begin{equation}
\label{vs}
\left\langle \left\{ v_{m,r}^{\prime\prime},
v_{m^{\prime},r^{\prime}}^{\dagger}\right\} \right\rangle 
=\frac 12\delta _{m+r,m^{\prime}+r^{\prime }}
\left\langle \Omega_{r,r^{\prime }}^{\prime\prime} \right\rangle ,
\end{equation}
where the precise order of the $X$-operators in
$\Omega_{r,r^{\prime}}^{\prime\prime}$ can be easily inferred from
(\ref{nu2pr}) and is given in the Appendix A. There, it is also shown that for 
slowly decaying spin correlation functions (as in the present case, see
(\ref{e25})) the correlation functions $\langle
\Omega_{r,r^{\prime}}^{\prime\prime} \rangle$ can be approximated to be
a function of $r-r^{\prime}$ only, in the way: 
\begin{equation}
\label{e37}
\langle \Omega_{r,r^{\prime}}^{\prime\prime}\rangle \approx 
\langle \Omega_{r-r^{\prime},0}^{\prime\prime}\rangle \approx 
\langle \Omega_{1}\rangle  \langle \Omega_{r-r^{\prime}}\rangle \; .  
\end{equation}
That leads to a constant shift of the energy $\epsilon_s$ as the only effect
of $v_{m,r}^{\prime\prime}$ which will be neglected further on. The
contribution of the $v_{m,r}^{\prime}$ term can be written analogously to
(\ref{vs}), defining the spin correlation functions $\langle
\Omega_{r,r^{\prime}}^{\prime} \rangle~=~\langle
\Omega_{r-r^{\prime},0}^{\prime} \rangle$.  
The correlation functions $\langle \Omega_{l,0}^{\prime}\rangle $ differ from 
$\langle \Omega_{l+1}\rangle $ only by the exchange
of two $X$-operators. Therefore, for
large $l$, we may expect that
\begin{equation}
\label{e38}
\langle \Omega_{l,0}^{\prime}\rangle \approx 
\langle \Omega_{l+1}\rangle \; . 
\end{equation}
That leads after 
Fourier transformation to the contribution of the $v_{m,r}^{\prime}$ term to
the spinon dispersion (Appendix A). Together with the contribution of 
$v_{k,q}$ we obtain for the spinon
dispersion
\begin{equation}
\label{e41}
\epsilon_s(Q-\pi )Z(Q)=\frac J2\left\{ \cos Q\left[ Z(Q)+\frac 12\right] + 
\frac 12\langle \Omega _{0,0}^{\prime }\rangle 
-\langle \Omega_{1} \rangle 
-\frac 12\sin Q\int_0^{2\pi }\frac{d\kappa }{2\pi }Z(\kappa
)\left( \cot \frac{Q-\kappa }2+\cot \frac{Q+\kappa }2\right) \right\} \; ,
\end{equation}
and the hole spectral function becomes
\begin{equation}
\label{AJ}A(k,\omega )=\int_{-\pi }^{+\pi }\frac{dQ}{\pi }Z(Q)\delta
\left[ \omega +2t\cos (k-Q)-2t_3\cos 2(k-Q)-\epsilon _s(Q-\pi )\right] .
\end{equation}
The curve that we obtained for $\epsilon _s$ with the formula (\ref{e41}) is
close to
\begin{equation}
\label{Es1}\epsilon _s(Q-\pi )\approx \alpha J\cos Q\quad ,\quad \alpha
\approx 2\;.
\end{equation}
as shown in Fig.\ 1. The functional form (\ref{Es1}) is consistent with
Bethe-ansatz \cite{SP} and field-theoretical considerations. \cite{Nagaosa}
(Sorella and Parola \cite{SP} derived a contribution $J\pi /2\cos Q\approx
1.6J\cos Q$.) We tested it also by comparing the first two terms of Fourier
expansion of the product $\cos QZ(Q)$ with $\langle \Omega_{0,0}^{\prime
}\rangle $ and $\langle \Omega_{1,0}^{\prime }\rangle$ 
that give values for $\alpha $ in (\ref{Es1}) of 2.1 or 1.8, respectively
(for the pair correlation functions we took the data of Ref.
\onlinecite{OgSh}). Therefore, we are using the simplified formula (\ref{Es1})
instead of (\ref{e41}) in the following analysis of the spectral density. 
We have checked that the differences are negligible.

The spectral density (for $t_3=0$) is shown in Fig.\ 2. One can clearly
distinguish 
between the spinon and holon features at the lower edge of the spectral
density dispersing at an energy scale $\propto J$ (from $k=0$ to $k=\pi/2$)
or $\propto t$ (from $k=\pi/2$ to $k=\pi$). 
At $k=k^*$, which is determined by $t \cos k^* = J$, another 
holon branch splits off the lower edge of the spectrum
\cite{Nagaosa} and disperses towards $k=0$ at an energy scale of $t$ (and a
corresponding holon branch splits off the upper edge of the spectrum).  
For $k$ values inbetween 0 and $k^*$ one has three peaks in the spectral
function (one 
spinon and lower and upper holon branch). 
One can easily imagine the situation in the doped case. Then the spinon and
holon branches start at the Fermi energy with two different velocities. 
\cite{Voit}  

In contrast to the 2D case, \cite
{MH} there is no separate bound state at the lower edge of the spectrum
indicating that there are only collective spin and charge
excitations. Most of those features were also observed in the ARPES experiment.
\cite{K96} In the naive picture of spin-charge separation (\ref{Astr}) the
spectral density would have square root singularities only either at the
lower or at the upper edge of the spectrum. In Fig.\ 2, however, there are
additional holon branches due to the square root singularity in $Z(Q)$ (see
also 
Ref.\ \onlinecite{Nagaosa}). For $J \to 0$ the spinon feature in the
spectral density, i.e.\ the lower edge of the spectrum, becomes
completely flat between $0$ and $\pi/2$. The corresponding pictures were
already given in Ref.\ \onlinecite{SP}. Fig.\ 2 agrees also qualitatively
with the finite cluster results. \cite{K96}

\section{Majumdar-Ghosh model}

We have shown that our approach is applicable for any
magnetic state for $J\to 0$. Now, we are going to present the spectral
function of one hole  
in the $t$-$J$-$J^{\prime}$ model with the special frustration
$J^{\prime}=J/2$ (called here Majumdar-Ghosh (MG) model for simplicity). In
that case rigorous analytic results may be obtained since the ground-state
wave function of the MG spin Hamiltonian is exactly known. \cite{MG} 
It is the combination of two simple dimer
states  
\begin{equation}
\label{e41a}
\Psi _{MG}=(\Phi _1+\Phi _2)/\sqrt{2} \; ,
\end{equation}
where 
$$
\Phi _1=\prod_{n=-\infty }^{+\infty }[2n,2n+1]\quad ,
\quad \Phi_2=\prod_{n=-\infty }^{+\infty }[2n-1,2n] \; ,
$$
and the singlet bond is denoted as 
$$
[l,m]\equiv \frac{1}{\sqrt{2}} 
\sum_\sigma \sigma X_l^{\sigma 0}X_m^{-\sigma 0}\left| vac\right\rangle \; .
$$
We are considering the MG model as a representative example for the case that
there is a gap in the spin excitation spectrum (and also in the charge
channel). To give the result for the spectral density in the strong coupling
limit $J \to 0$ one has to find the modified quasiparticle residue $Z(Q)$
in (\ref{AJ0}). It can be simply derived from the correlation
functions (see also Ref.\ \onlinecite{Maekawa})
\begin{eqnarray}
\langle \Omega_{l} \rangle &=&
\frac 12 \left[ \langle \Phi_1 | \Omega_{l} | \Phi_1 \rangle
+ \langle \Phi_2 | \Omega_{l} | \Phi_2 \rangle \right]
\nonumber \\
\langle \Omega_{2n} \rangle &=&
\left(- \frac 12 \right)^n
\quad , \quad
\langle \Omega_{2n+1} \rangle =
\frac 12 \left( - \frac 12 \right)^{n+1} 
\quad , \quad n \geq 0 \; ,
\label{f1}
\end{eqnarray}
in the following way 
\begin{equation}
\label{45}
Z(Q)=\frac 12+\sum_{n=1}^\infty \left[ \left( \frac 12-\frac{{\rm e}^{-iQ}}4 
\right) \left( -\frac{{\rm e}^{2iQ}}2\right) ^n+h.c.\right] =\frac 32\frac{ 
1+\cos Q}{5+4\cos 2Q} \; . 
\end{equation}
The corrections for small $J \ll t$ may only be derived approximatively and we 
present two methods, projection method and variational procedure having 
different accuracy. 

\subsection{Projection method}

First we calculate the spinon dispersion $\epsilon_s$ in the same way as it
was done in the Heisenberg case in Sec.\ IV. But we should keep in mind that
its applicability is less justified for the MG model than for the pure $t$-$J$ 
model due to the much faster decay of spin correlation functions (compare
(\ref{f1}) with (\ref{e25})). As before, we approximate $\langle
\Omega_{r,r^{\prime}}^{\prime\prime}\rangle \approx \langle \Omega_1 \rangle
\langle \Omega_{r-r^{\prime}} \rangle$ which results in a constant energy
shift from the $v_{k,q}^{\prime\prime}$ term. Therefore, the first
contribution to $\epsilon_s$ coming from $\hat{J}$ is merely determined by
$\langle \Omega_{r,r^{\prime}}^{\prime} \rangle$ (see Appendix B) resulting in
\begin{equation}
\epsilon _{sJ}(Q-\pi )=2J\cos Q \, . 
\end{equation}
We have a second contribution to $\epsilon_s$ from 
$\hat J^{\prime }$
\begin{equation}
\label{fe}
\epsilon _{sJ^{\prime }}(Q-\pi )=J^{\prime }\left[ -4\cos Q+\frac 5 
4+\cos 2Q\right] \;.
\end{equation}
We see that for $J^{\prime }=J/2$ the terms proportional $\cos Q$ cancel and
we find
\begin{equation}
\label{sdf}\epsilon _s (Q - \pi) =J\left[ \frac 58+\frac 12\cos 2Q\right] \;,
\end{equation}
which is symmetric around $\pi /2$.  

The spectral density is presented in Fig.\ 3. We see that in contrast to the
$t$-$J$ model the structures coming from $Z(Q)$ (the holon branches) are much
less pronounced, whereas square root singularities exist at the lower and upper
edges of the spectrum. Their intensities are proportional to $Z(k)$ or
$Z(k-\pi)$ at the lower and upper edges, respectively. Therefore, the square 
root singularity vanishes for $k=\pi$ at the lower edge. Furthermore, one
can see that the low energy region for $k$ between $\pi/2$ and $\pi$ being 
empty in Fig.\ 2 is now filled with states. The spectrum becomes more
symmetric around $\pi/2$ and the low-energy edge is given by the spinon
dispersion $\epsilon_s(Q)$. The strong damping of the holon branch
is due to the suppression of the singularity at the spinon Fermi edge (at
$Q=\pi/2$ in $Z(Q)$). It is a universal feature for any 1D magnetic state
having a gap in the spin excitation spectrum. The suppression of holon weight
was also found by Voit \cite{Voit2} for the Luther-Emery phase in the
Luttinger liquid. 
The form of the spectral density in Fig.\ 3 
resembles also roughly the exact diagonalization study in Ref.\ 
\onlinecite{Maekawa}. But a single bound state with a finite spectral weight
that was obtained there, is missing in Fig.\ 3. That deficiency is due to the
special projection procedure (\ref{ptdec}) which can only result in a
continuous spectral density. Therefore, one has to go beyond the projection
method.

\subsection{Variational ansatz}

Here we will use the set of string operators (\ref{vmr})
as a set defining a variational wave function for the whole Hamiltonian. 
Due to the knowledge of the exact ground-state (\ref{e41a}) all necessary
matrix elements can be calculated without any further approximation. 
More precisely, we will diagonalize the 
Hamiltonian $\hat H = \hat t + \hat J + \hat J^{\prime}$ 
in the space spanned by the set of basis 
operators 
\begin{equation}
v_{k,r}=\frac{1}{\sqrt{L}} \sum_{-\infty}^{+\infty} \mbox{e}^{i k (m+r)}
v_{m,r} \; \; ,
\end{equation}
where $L$ is the number of lattice sites and $v_{m,r}$ was defined in (18). 
For that purpose one has to calculate the overlap matrix resulting in
\begin{equation}
S_{r,r^{\prime}}=\left\langle \{v_{k,r},v_{k,r^{\prime}}^{\dagger} \} \right
\rangle = \frac{1}{2} \langle \Omega_{r-r^{\prime}} \rangle \; .
\end{equation}
The kinetic energy part of the Hamilton matrix is given by:
\begin{equation}
\frac{t}{2} E_{r,r^{\prime}}^k =
\left \langle \left\{ [ v_{k,r}, \hat t ] , 
v_{k,r^{\prime}}^{\dagger} \right\} \right
\rangle =
\frac{t}{2} \left( \mbox{e}^{i k} \langle \Omega_{r-r^{\prime}-1} \rangle 
+ \mbox{e}^{- i k} \langle \Omega_{r-r^{\prime}+1} \rangle \right) \; . 
\end{equation}
The calculation of the exchange part of the Hamilton matrix is quite lengthy
but straightforward. It shall not be given here in detail. To present the
results we define a matrix $\underline{\underline E}^x$:
\begin{equation}
\frac{J}{4} E_{r,r^{\prime}}^x=
\left \langle \left\{ [ v_{k,r}, \hat J + \hat{J^{\prime}} ] ,
v_{k,r^{\prime}}^{\dagger} \right\} \right \rangle \; , 
\end{equation}
whose matrix elements are listed in Appendix C. 
The Hamilton matrix is then given by
\begin{equation}
\underline{\underline E}=
\frac{t}{2} \underline{\underline E}^k +
\frac{J}{4} \underline{\underline E}^x \; , 
\end{equation}
and the matrix GF
\begin{equation}
G_{r,r^{\prime}}
= \langle\langle v_{k,r}|v_{k,r^{\prime}}^{\dagger}\rangle\rangle
\end{equation}
can be found by solving the equation
\begin{equation}
(\omega+i\Gamma+ - \underline{\underline E} \underline{\underline S}^{-1} ) 
\underline{\underline G} =  
\underline{\underline S} 
\; , \quad \Gamma > 0 
\; . 
\end{equation}
Finally, the GF (\ref{rGf}) can be obtained by $G(k,\omega)=2G_{0,0}$. 

The numerical results for $J=0$ and $J=0.4$ at three different momenta are
presented in Fig.\ 4. The curves for $J=0$ 
coincide with the analytic expression (\ref{AJ0},\ref{45}). A number of 400
basis functions and a broadening of $\Gamma=0.05$ are sufficient to reach the
thermodynamic limit in contrast to the exact diagonalization method yielding
only a sequence of $\delta$-peaks. For $J=0.4$ we can confirm the features
found by the projection method, i.e.\ the low energy intensity between $\pi/2$ 
and $\pi$, the symmetric spinon dispersion and the overdamped holon
branch. In addition, the exchange terms produce two new features not present
in Sec.\ VA: a resonance peak near zero energy and a bound state below 
the continuum. The resonance peak is visible near $k=\pi$ and becomes an
antiresonance near $k=0$. Careful inspection of the exact diagonalization
data \cite{Maekawa} indicates also a very high peak at the resonance position
for $k=\pi$ and a small gap at $k=0$, but a better understanding of the
resonance/antiresonance feature is still required.

The bound state is not visible in Fig.\ 4 due to the broadening $\Gamma$ 
which is too large. Instead, we present in Fig.~5 the spectral weight of the
lowest 
eigenstate $w_1$ for $k=\pi/2$ and $J=0.4$ in dependence on the number of
basis functions. It is clearly seen that the weight tends to a constant value
($w_1\approx 0.1$) in difference to the weight $w_3$ of the third
eigenstate. \cite{remark} 
At the same time, the separation $e_1=E_3-E_1$ between the first and the third 
eigenvalues $E_{1/3}$ stays finite for $N\to \infty$ but the separation is
very small 
($e_1 \approx 0.02$ in units of $t$). For $J=0$, both $w_1$ and $e_1$ tend to
zero for $N \to \infty$. That means that the bound state is connected with the 
presence of a gap in the spin excitation spectrum.

\section{Ideal paramagnetic state}

Such a state is realized for very high temperatures $T$, much larger than the
exchange energy $k_B T \gg J$. In that case spins at neighboring sites are
completely uncorrelated. But the temperature is assumed to be lower than the
Hubbard $U$ such that the constraint of no double occupancy is preserved. Then 
the correlation functions become simply
$$
\langle \Omega_{l} \rangle = \left( \frac 12 \right)^l \; ,
$$
which results in
\begin{equation}
\label{52}
Z(Q)=\frac 38 \frac{1}{\frac 54 + \cos Q} \; .
\end{equation}
The calculation of the spinon part (without frustration) gives
\begin{equation}
\label{st}\epsilon_s (Q-\pi) = \frac J2 \left[ 2 \cos Q + \frac 12 \right]
\; .
\end{equation}
To calculate it one has to note that (\ref{e38}) is no approximation in the
present case. The effect of the $\langle \Omega_{r,r^{\prime}}^{\prime\prime}
\rangle$ terms
can only be treated approximatively (see (\ref{e37})) but it was checked by
the variational method that its influence on the spectral function can be
neglected. 

The information on $Z(Q)$ and $\epsilon_s$ is sufficient to calculate the
spectral function (Fig.\ 6). 
It is surprising that the strong singularities at the band edges survive
despite the large temperature. The lower edge disperses according to the
dispersion of the spinon (\ref{st}) with a width proportional to $J$ and
has its minimum at $k=\pi$ (in contrast to the frustrated case Fig.\ 3 with 
a minimum at $k=\pi/2$). But a
peak connected with the holon dispersion proportional to $t$ is not seen in
Fig.\ 6. Such a peak appears in the finite temperature spectral function of
the 2D $t$-$J$ model \cite{VB} and it can be expected since the first
moment of the spectral function disperses according to $t \cos k$. Its
absence in 1D is a nontrivial and unexpected result.
It can be understood in the present context since the holon branch is strongly 
damped due to the suppression of the singularity in $Z(Q)$ at
$Q=\pi/2$. Apparently, that suppression is more strong in (\ref{52}) than in
the frustrated case (\ref{45}) such that the holon branch is still visible in
Fig.\ 3 but it disappears nearly in Fig.\ 6. One should note that the above
result holds only in the region $J \ll k_B T \ll U$. One may speculate that a
further increase of the temperature such that the constraint of no double 
occupancy is lifted should lead to drastic changes in the spectral
function. The strong singularities at the lower or upper band edges should
disappear and a free dispersion should become visible.

\section{Conclusion}

In conclusion we could derive analytic expressions for the spectral function
of one hole in several magnetic states. The expressions are rigorous in the
limit $J \to 0$, but our approach allows also to calculate the small $J$
corrections. We 
analyzed the frustration and temperature effects. Results were given for
the special frustration $J^{\prime}=J/2$ with 
a gap in the spin excitation spectrum
and for the ideal paramagnetic case. Both
effects, frustration and temperature, lead to low-energy excitations between
$\pi/2$ and $\pi$, and to a strong damping of the holon branches in the
spectral function caused by the suppression of the singularity at the Fermi
edge of spinons. 
The exchange terms in the MG model were found to be responsible for the finite 
weight of the lowest eigenstate and its finite, but small, energy separation
from the rest of the spectrum, i.e.\ the bound state. 
The proposed scenario of holon branch damping seems to be a universal feature
of frustration and temperature. Therefore, our results are of direct
importance for photoemission experiments on strongly frustrated 1D compounds
like CuGeO$_3$, for instance. However, edge-shared cuprate chains have a
smaller energy scale and less ideal 1D behavior in comparison with
corner-shared compounds, \cite{RoDre} which hinders
direct comparison with experiment. But it cannot be excluded that a small
frustration is also present in SrCuO$_2$ such that our study gives one
possible reason, why no real, separate holon
branch could be observed in the experimental spectra of SrCuO$_2$ between
$k=0$ and   
$\pi/2$. \cite{K96} In the spin gap case we found
a very small energy separation of the bound state from the continuum such that 
it is nearly impossible to detect it in a photoemission experiment.

\vspace*{1cm}

{\bf Acknowledgements}

\vspace*{1cm}

The authors thank the DFG, the INTAS organization
(project No.\ INTAS-97-11066), and the Max-Planck society for financial
support and S.-L.\ Drechsler, W.\ Brenig, K.\ Becker, A.\ Muramatsu and H.\
Eschrig for discussions. R.K.\ acknowledges the
hospitality of the University of Technology Dresden, where the main part of
this work has been carried out. R.K.\ also thanks O.A.\ Starykh for valuable
discussions and providing the water-way in the ocean of literature on 1D
magnetism.

\newpage
\appendix
\setcounter {section} {1}
\begin{center}
{\Large\bf Appendix A: Spinon dispersion of the Heisenberg case}
\end{center}
\newcounter {pzz}
\setcounter{equation}{0}
\setcounter{pzz}{\value{equation}}
\renewcommand {\theequation}{\Alph{section}.\arabic{equation}} 

In this Appendix we outline the main steps to derive the spinon dispersion of
the pure $t$-$J$ model using the projection method. 
For long chains of $X$-operators it is convenient to introduce the 
notations \cite{AFB} 
$$
\sum_{\alpha _1,\ldots ,\alpha _r}X_{n_1}^{\sigma \alpha _1}X_{n_2}^{\alpha
_1\alpha _2}X_{n_3}^{\alpha _2\alpha _3}\ldots X_{n_{r-1}}^{\alpha
_{r-1}\alpha _r}X_{n_r}^{\alpha _r0}\equiv (n_1|n_2|n_3|\ldots |n_{r-1}|n_r]
$$
and
$$
\sum_{\sigma ,\alpha _1,\ldots ,\alpha _r}X_{n_1}^{\sigma \alpha
_1}X_{n_2}^{\alpha _1\alpha _2}X_{n_3}^{\alpha _2\alpha _3}\ldots
X_{n_{r-1}}^{\alpha _{r-1}\alpha _r}X_{n_r}^{\alpha _r\sigma }\equiv
(n_1|n_2|n_3|\ldots |n_{r-1}|n_r)\;.
$$
which means especially that (\ref{vmr}) may be rewritten as
$v_{m,r}\equiv (m|m+g|m+2g|\ldots |m+r]$.
In such a notation we obtain for the commutation with the Heisenberg
Hamiltonian 
\begin{equation}
\label{Xs0J}\left[ X_m^{\sigma 0},\hat J\right] =-\frac J2\sum_{g,\gamma
}X_{m+g}^{\sigma \gamma }X_m^{\gamma 0}=-\frac J2\sum_g(m+g|m] \; ,
\end{equation}
\begin{equation}
\label{XabJ}\left[ X_m^{\alpha \beta },\hat J\right] =\frac J2\sum_{g,\gamma
}\left( X_m^{\alpha \gamma }X_{m+g}^{\gamma \beta }-X_{m+g}^{\alpha \gamma
}X_m^{\gamma \beta }\right) =\frac J2\sum_g\{(m|m+g|-(m+g|m|\} \; ,
\end{equation}
and then 
\begin{eqnarray}
\left[ v_{m,r},\hat J\right] &=& \frac J2
\left\{ (m|m-g|m+g|m+2g|\ldots|m+r]
-(m-g|m|m+g|m+2g|\ldots |m+r]
\right.
\nonumber \\
&&
\left.
-(m+g|m+2g|\ldots |m+r]
-(m|\ldots |m+r-g|m+r+g|m+r]\right\} \; ,
\nonumber \\
\label{comJapp}
\end{eqnarray}
where $g={\rm sign}(r)$. In deriving (\ref{comJapp}) it is important that the
commutation of the ''inner'' operators 
$X_{m+l}^{\alpha \beta }$ with $l<r$ do not give rise to additional terms
since the corresponding sums cancel each other. That is a direct consequence
of one-dimensionality.

Now, we consider the holon contribution to $\epsilon_s$ coming from
$v_{k,q}^{\prime\prime}$
\begin{equation}
\label{com2prim}
\left\langle \left\{ 
v_{k,q}^{\prime \prime },v_{k^{\prime },q^{\prime}}^{\dagger}
\right\} \right\rangle 
=2\pi \delta (k-k^{\prime })\sum_{r,r^{\prime }}
\mbox{e}^{iq^{\prime}r^{\prime}-iqr+ik(r-r^{\prime})}
\langle \Omega_{r,r^{\prime}}^{\prime\prime}\rangle
\end{equation}
with
\begin{equation}
\label{a6}
\Omega_{r,r^{\prime}}^{\prime \prime }=
(m|\ldots |m+r-g|m+r+g|m+r|m+r-g^{\prime}|\ldots |m+r-r^{\prime }) \; ,
\end{equation}
and $g=\mbox{sign}(r)$, $g^{\prime}=\mbox{sign}(r^{\prime})$. We see that in
general $\langle \Omega_{r,r^{\prime}}^{\prime\prime}\rangle$ depends both on
$r-r^{\prime}$ and on $r$. Eqn.\ (\ref{com2prim}) can also be written as 
\begin{equation}
\left\langle \left\{ 
v_{k,q}^{\prime \prime },v_{k^{\prime },q^{\prime}}^{\dagger}\right\} 
\right\rangle 
=2\pi \delta (k-k^{\prime })\sum_{l}
\mbox{e}^{i(k-q)l} S_l \; ,
\end{equation}
with
$$
S_l=\sum_{r=-\infty}^{+\infty} \mbox{e}^{-i(q-q^{\prime})r}
\langle \Omega_{r,r-l}^{\prime\prime}\rangle \; .
$$
Due to the slow decay of spin correlation functions in the 1D Heisenberg
state, 
one can expect that the main contribution to $S_l$ comes from regions where
$|r|\gg |l|$. There holds $g=g^{\prime}$ and we may rewrite and
approximate (\ref{a6}) by 
\begin{equation}
\langle \Omega_{r,r-l}^{\prime\prime}\rangle
= \langle (0|\ldots|l)(r+g|r)\rangle
\approx \langle \Omega_l \rangle \langle \Omega_1 \rangle \; .
\end{equation}
Then, the explicit dependence on $r$ drops out and we obtain
\begin{equation}
\left\langle \left\{ 
v_{k,q}^{\prime \prime },v_{k^{\prime },q^{\prime}}^{\dagger}\right\} 
\right\rangle 
= 8 \pi^2 \delta(k-k^{\prime}) \delta(q-q^{\prime}) Z(k-q+\pi)
\langle \Omega_1 \rangle \; ,
\end{equation}
i.e.\ a simple constant shift of the energy $\epsilon_s$.

The contribution of the $v_{k,q}^{\prime}$ term to the spinon dispersion is
determined by the sequence of spin operators 
\begin{equation}
\Omega_{r,r^{\prime}}^{\prime} =
(m|m-g|m+g|\ldots|m+r|m+r-g^{\prime}|\ldots|m+r-r^{\prime})
\end{equation}
instead of (\ref{a6}). The expectation value of that term has to be calculated 
for the magnetic system without holes. In difference to $\langle
\Omega_{r,r^{\prime}}^{\prime\prime}\rangle$, it depends only on
$r-r^{\prime}$  
without further approximation
\begin{equation}
\langle \Omega_{r+l,r}^{\prime}\rangle = 
\langle \Omega_{l,0}^{\prime}\rangle = 
\langle (0|-1|1|\ldots|l)\rangle \; ,
\quad
(l>0) \; ,
\end{equation}
and $\langle \Omega_{-l,0}^{\prime}\rangle = 
\langle \Omega_{l,0}^{\prime}\rangle$. For $l=0, 1$ it can
be expressed through pair correlation functions
\begin{eqnarray}
\langle \Omega_{0,0}^{\prime} \rangle &=&
\frac{1}{2} + 2 \langle {\bf S}_{0} {\bf S}_{2} \rangle \; , 
\nonumber \\
\langle \Omega_{1,0}^{\prime} \rangle &=&
\frac{1}{4} + 2 \langle {\bf S}_{0} {\bf S}_{1} \rangle
+ \langle {\bf S}_{0} {\bf S}_{2} \rangle \; . 
\nonumber
\end{eqnarray}
For large $l>0$ we may expect
\begin{equation}
\langle \Omega_{l,0}^{\prime}\rangle \approx \langle \Omega_{l+1}\rangle 
\; .
\end{equation}
Using this approximation we obtain the following contribution to the spinon
dispersion $\epsilon_s$ which stems from the $v_{k,q}^{\prime}$ term
\begin{equation}
\frac{J}{2} \left\langle \left\{ 
v_{k,q}^{\prime},v_{k^{\prime },q^{\prime}}^{\dagger}
\right\} \right\rangle 
=8\pi^2 \delta(q-q^{\prime}) \delta (k-k^{\prime })
\epsilon_s^{\prime}(k-q)Z(k-q+\pi)
\end{equation}
with
\begin{equation}
\epsilon_s^{\prime}(k)Z(k+\pi)
= \frac{J}{4} \sum_{l=-\infty}^{+\infty}
\mbox{e}^{-ikl}
\langle \Omega_{l,0}^{\prime}\rangle \; .
\end{equation}
After some algebra we find
\begin{equation}
4 \epsilon_s^{\prime} (Q-\pi)Z(Q)/J =
\langle \Omega_{0,0}^{\prime} \rangle
- 2 \langle \Omega_{1} \rangle
- 2 \cos Q \left[ Z(Q) - \frac 12 \right] - \sin Q Y(Q) \; , 
\label{App}
\end{equation}
where
\begin{equation}
Y(Q) = \sum_{l=1}^{+\infty} 2 (-1)^l \sin (Ql)
\langle \Omega_{l} \rangle
= \frac{1}{2\pi} \int_{0}^{2 \pi} d \kappa Z(\kappa)
\left[ \cot \frac{Q-\kappa}{2} + \cot \frac{Q+\kappa}{2} \right] \; .
\nonumber
\end{equation}
The complete expression for the spinon dispersion follows from 
$$
\epsilon_s(Q-\pi )= J \cos Q + \epsilon _s^{\prime }(Q-\pi) +{\rm const}
\; , 
$$
and is given in Eqn.\ (\ref{e41}) neglecting the constant energy shift. 

\newpage
\appendix
\setcounter {section} {2}
\begin{center}
{\Large\bf Appendix B: Majumdar-Ghosh model with projection method}
\end{center}
\newcounter {pzzb}
\setcounter{equation}{0}
\setcounter{pzzb}{\value{equation}}
\renewcommand {\theequation}{\Alph{section}.\arabic{equation}} 

The commutation with the frustration Hamiltonian (\ref{Jprime}) is very
similar to 
(\ref{comJapp}). It gives 
\begin{equation}
\label{comJpr}\left[ v_{m,r},\hat J^{\prime }\right] =\frac{J^{\prime }}2 
\left\{ v_{m,r}^{(3)}+v_{m,r}^{(4)}\right\}
\end{equation}
where
\begin{eqnarray}
v_{m,r}^{(3)}&=&
\left\{(m|m-2g|m+g|\ldots|m+r]-
(m-2g|m|m+g|\ldots|m+r] \right.
\nonumber \\
&&-(m|m+g)(m+2g|\ldots|m+r]
+(m|m+g|m-g|m+2g|\ldots|m+r]
\nonumber \\
&&-\left.(m|m-g|m+g|\ldots|m+r]\right\} \; , 
\nonumber \\
v_{m,r}^{(4)}&=&
\left\{-(m|\ldots|m+r-g|m+r+2g|m+r]\right.
\nonumber \\
&&-(m|\ldots|m+r-2g|m+r+g|m+r-g|m+r]
\nonumber \\
&&+\left.(m|\ldots|m+r-g|m+r+g|m+r]\right\} \; . 
\end{eqnarray}
Again we see that all terms coming from commutations at ``inner''
operators cancel. But now the motion of spinons becomes more
complicated. The same considerations that show the absence of dispersion
from the $v_{m,r}^{\prime \prime }$ term are applicable to $v_{m,r}^{(4)}$. 

Taking into account that
$$
\left\langle \Omega _{0,0}^{\prime }\right\rangle =\frac 12 
\; , \;
\left\langle \Omega _{2n,0}^{\prime }\right\rangle =-\frac 14 
\left( -\frac 12\right)^n \; , \; 
\left\langle \Omega _{2n-1,0}^{\prime}\right\rangle 
=\left( -\frac 12\right)^n \; , \; n>0 \; , 
$$
we obtain the following contributions to $\epsilon _s=\epsilon
_{sJ}+\epsilon _{sJ^{\prime }}$ (we drop dispersionless terms). 
From $\hat J$ comes
\begin{eqnarray}
\epsilon_{sJ}(Q-\pi)Z(Q) &=& \frac J2 \left[
2 \cos Q Z(Q) + Z^{\prime}(Q) \right] \; , 
\nonumber \\
Z^{\prime} (Q) &=& \frac{3(-\frac 18 + \cos Q)}{5+4 \cos 2Q} + \frac 14
+ \frac 18
=Z(Q) 2 \cos Q \; . 
\end{eqnarray}
Thus
\begin{equation}
\epsilon _{sJ}(Q-\pi )=2J\cos Q
\end{equation}
has the same form (\ref{Es1}) that we have assumed for the $t$-$J$ model.

For the term that comes from the left distorted end of $v_{m,r}$ due to
$\hat J^{\prime}$ we have
$$
\left\langle \left\{ v_{m,r_1}^{(3)},v_{m+r_1-r_2,r_2}^{\dagger }\right\}
\right\rangle \equiv \frac{1}{2} \left\langle \Omega _{r_1-r_2}^{(3)}
\right\rangle 
\; , 
$$
$$
\left\langle \Omega _{2n}^{(3)}\right\rangle =\left( -\frac 12 
\right)^n \; , \; 
\left\langle \Omega _{2n-1}^{(3)}\right\rangle
=-4\left( -\frac 12\right)^n \; , \; n>1 \; , 
$$
$$
\left\langle \Omega _{0}^{(3)}\right\rangle =\left( -\frac 12 
\right) \; , \; 
\left\langle \Omega_{1}^{(3)}\right\rangle
=\frac 54 \; . 
$$
The contribution from $\hat J^{\prime }$ is 
\begin{eqnarray}
\epsilon_{sJ^{\prime}}(Q-\pi)Z(Q) &=& \frac {J^{\prime}}{2} \left[
- 4 Z^{\prime}(Q)  + \frac 34 + \frac 68  \cos Q \right]
\nonumber \\
&=& Z(Q) \frac{J^{\prime}}{2} \left[
- 8 \cos Q + 2 ( \frac 54 + \cos 2Q)\right]
\end{eqnarray}
and we obtain (\ref{fe}).

\newpage
\appendix
\setcounter {section} {3}
\begin{center}
{\Large\bf Appendix C: Matrix elements of the variational basis set}
\end{center}
\newcounter {pzzc}
\setcounter{equation}{0}
\setcounter{pzzc}{\value{equation}}
\renewcommand {\theequation}{\Alph{section}.\arabic{equation}}

The matrixelements of $E_{r,r^{\prime}}^{x}$ in the neighborhood of
$r,r^{\prime}=0$ are given by: 

\begin{center}
\begin{tabular}{|c|c||c|c|c|c|c|c|c||}
\hline
 & $r^{\prime}$ & -3 & -2 & -1 & 0 & 1 & 2 & 3 \\ \hline
$r$&            & & & & & & & \\ \hline \hline
-3 & & 3/8 & 0 & 3/16 & -3/32 & -9/32 & 3/32 & 3/16 \\ \hline   
-2 & & 0  & 3/8 & -9/16 & 3/8 & 0 & -3/8 & 3/32 \\ \hline      
-1 & & 3/16  & -9/16 & 3/8 & -3/16 & 3/8 & 0 & -9/32 \\ \hline      
 0 & & -3/32 & 3/8 & -3/16 & 0 & -3/16 & 3/8 & -3/32 \\ \hline   
 1 & & -9/32 & 0 & 3/8 & -3/16 & 3/8 & -9/16 & 3/16 \\ \hline
 2 & &  3/32 & -3/8 & 0 & 3/8 & -9/16 & 3/8 & 0 \\ \hline
 3 & & 3/16 & 3/32 & -9/32 & -3/32 & 3/16 & 0 & 3/8 \\ \hline \hline
\end{tabular}
\end{center}
One has two different regions in the matrix. The first one is defined for
$r>0, \; r^{\prime}>0$ and $r \geq r^{\prime}+2$ where we have the matrix
elements:
\begin{equation}
\begin{array}{ll}
r=2n, \;  r^{\prime}=2m \quad \mbox{or}  
\quad r=2n+1, \; r^{\prime}=2m+1 : & E_{r,r^{\prime}}^{x} =
- \frac{3}{8}  \left( - \frac{1}{2} \right)^{n-m} \; , \\
r=2n+1, \; r^{\prime}=2m:  & E_{r,r^{\prime}}^{x} =
 \frac{3}{8} \left( - \frac{1}{2} \right)^{n-m}  \; , \\ 
r=2n, \; r^{\prime}=2m-1: & E_{r,r^{\prime}}^{x} =
- \frac{3}{16}  \left( - \frac{1}{2} \right)^{n-m+1} \; ,\\ 
\end{array}
\end{equation}
and the second one for $r^{\prime} \leq -2, \; r \geq 2$ with
\begin{equation}
E_{r, r^{\prime}}^{x} = - \frac{3}{2} 
\langle \Omega_{r-r^{\prime}}  \rangle  \; .
\end{equation}
There are special matrix elements along the diagonal (3/8) and along the side
diagonal (alternatively -9/16 or 0) and also for the two lines $(n \geq 1)$:
\begin{eqnarray*}
E_{2n,0}^{x} = - \frac{3}{4} \left( - \frac{1}{2} \right)^n 
\quad 
&& \quad E_{2n+1,0}^{x} =  \frac{3}{16} \left( - \frac{1}{2} \right)^n
\\   E_{2n,-1}^{x} = 0 \quad && \quad  
 E_{2n+1,-1}^{x} =  \frac{9}{16} \left( - \frac{1}{2} \right)^n   
\end{eqnarray*}
The matrix is filled by
\begin{equation}
E_{-r,-r^{\prime}}^{x}=         
E_{r,r^{\prime}}^{x}=         
E_{r^{\prime},r}^{x} \; .
\end{equation}

\begin{center}
{\Large {\bf Figures}}
\end{center}

Fig.\ 1: Comparison of the spinon dispersion $\epsilon_s(Q-\pi)$ as calculated
from the projection method (broken line, $J=0.4$) with $2J\cos Q$ (full
line). 

Fig.\ 2: Spectral density of the $t$-$J$ model for $J=0.4$ and $t=1$.

Fig.\ 3: Spectral density of the frustrated $t$-$J$ model ($J=0.4$ and $t=1$)
at the special frustration $J^{\prime}=0.5 J$ (using the Majumdar-Ghosh wave
function) within the projection method. 

Fig.\ 4: Spectral density of the Majumdar-Ghosh model $A(k,\omega)$ for three
different momenta $k/\pi$ 
and $t=1$, $J=0.4$ (full lines) or $J=0$ (dashed lines) with a variational
set of 400 basis functions and a broadening of $\Gamma=0.05$. 

Fig.\ 5: Weight $w_1$ and energy separation $e_1=E_3-E_1$ of the lowest
eigenvalue $E_1$ at $k/\pi=0.5$, $t=1$, $J=0.4$ (full lines) as a function of
the inverse number of basis functions $1/N$. The dashed lines are the weights
$w_3$ and the energy separation $e_3=E_5-E_3$ of the third eigenvalue $E_3$.  

Fig.\ 6: Spectral density of the $t$-$J$ model ($J=0.4$ and $t=1$) in the
ideal paramagnetic state. 

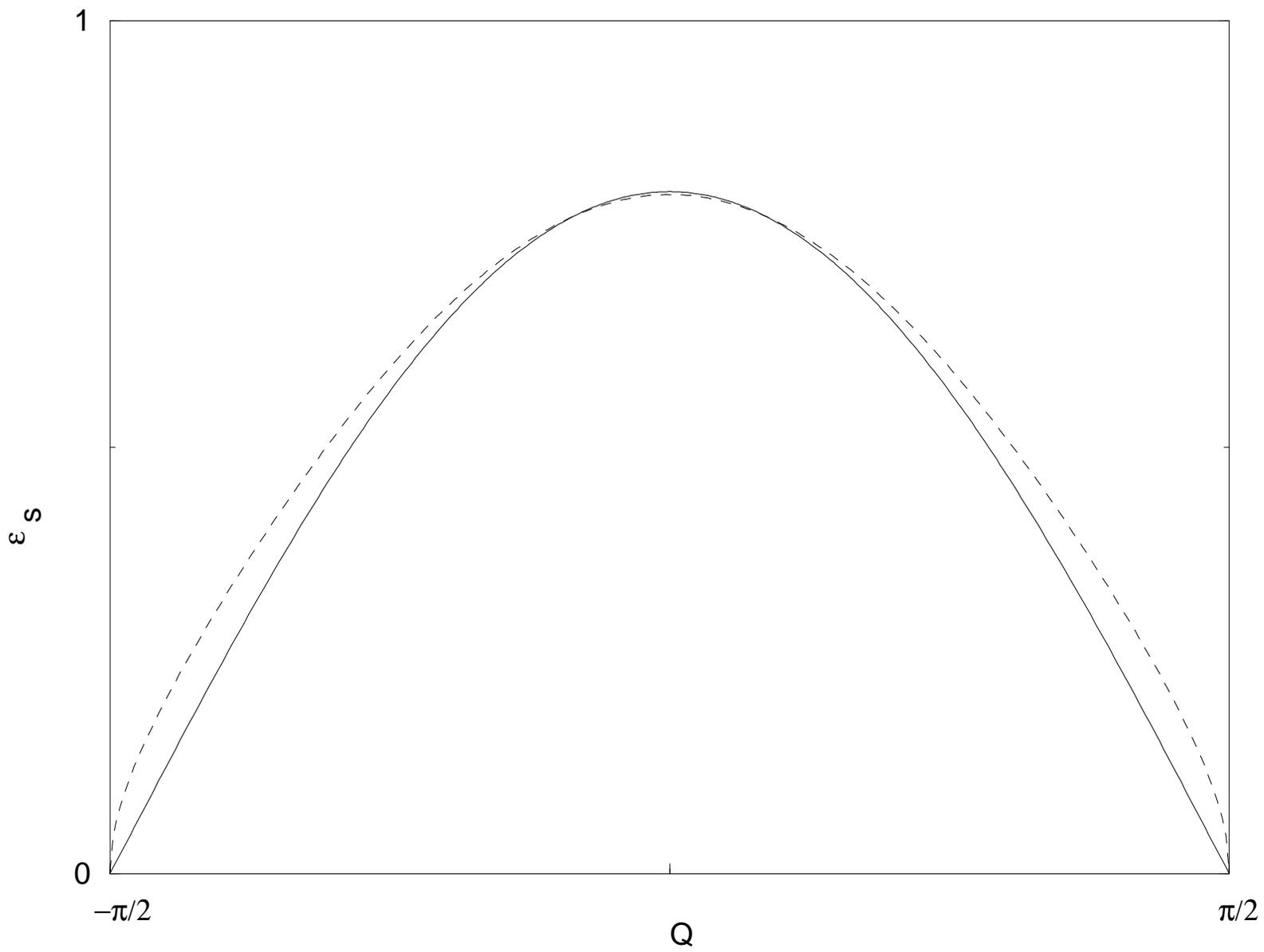
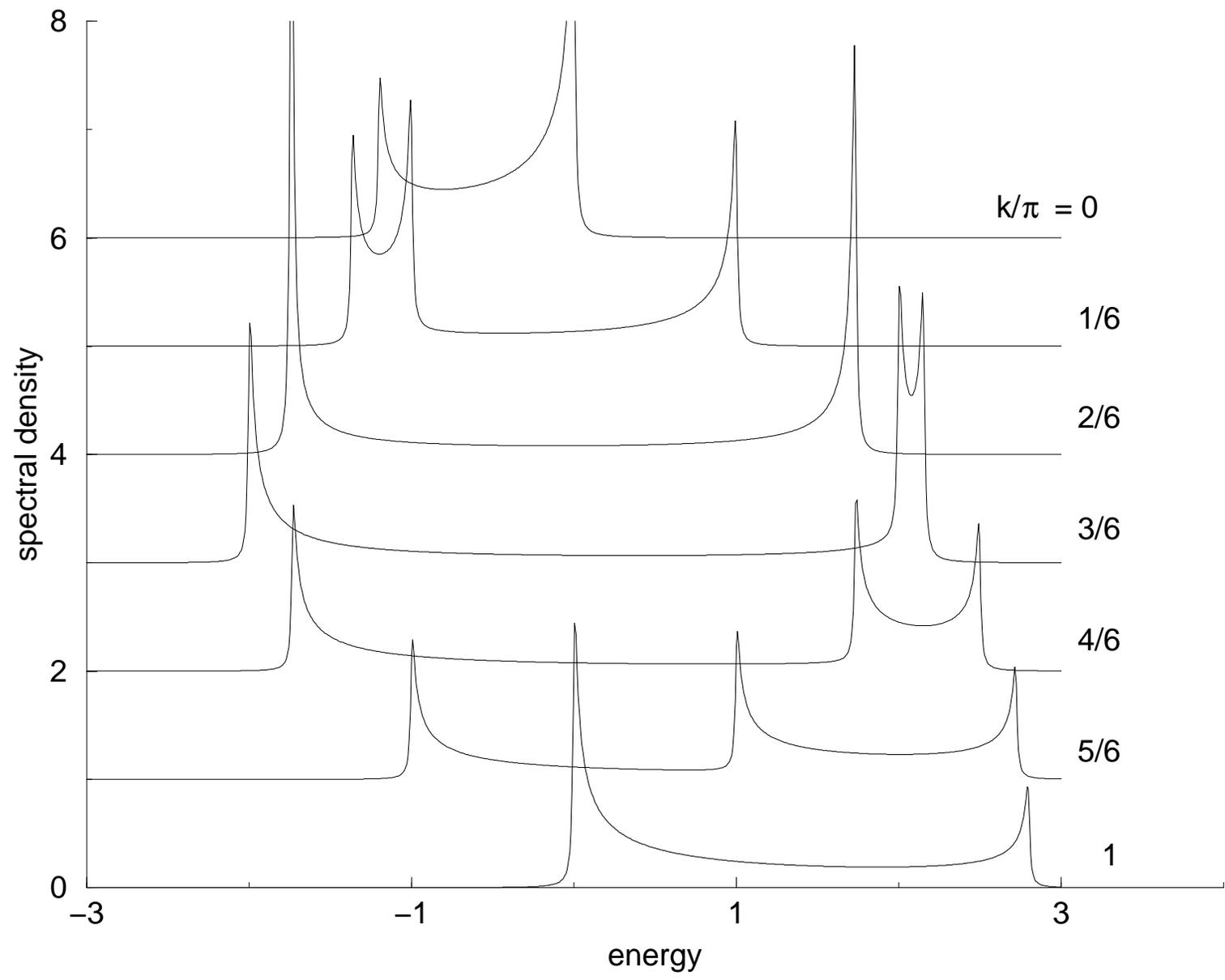
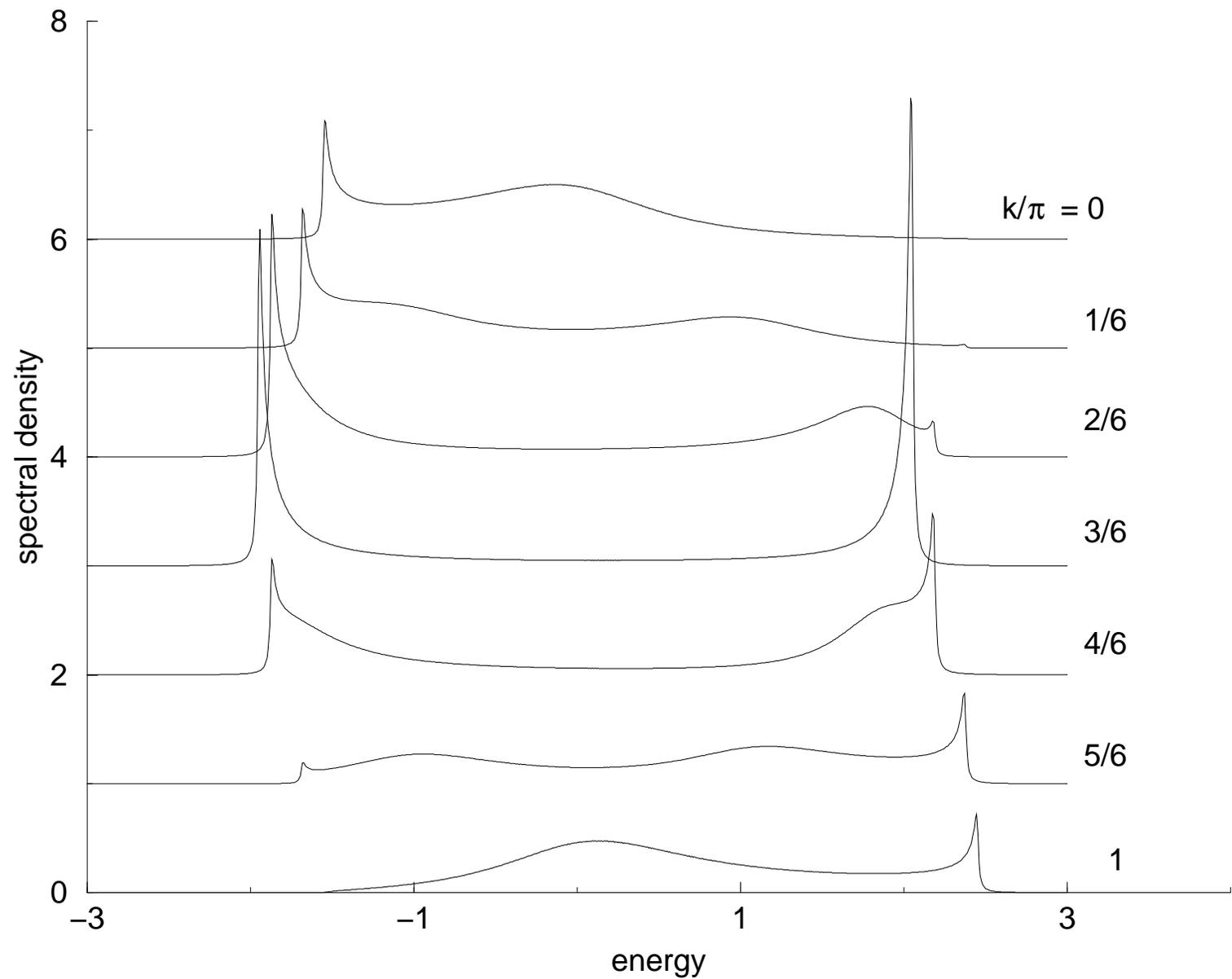
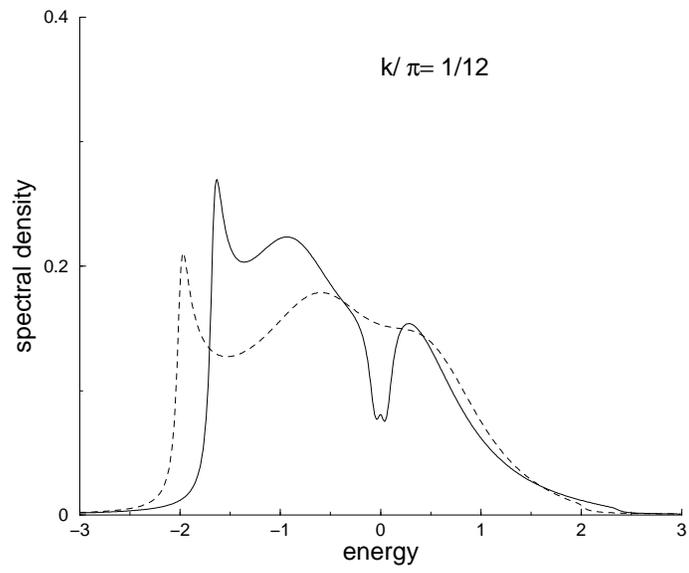
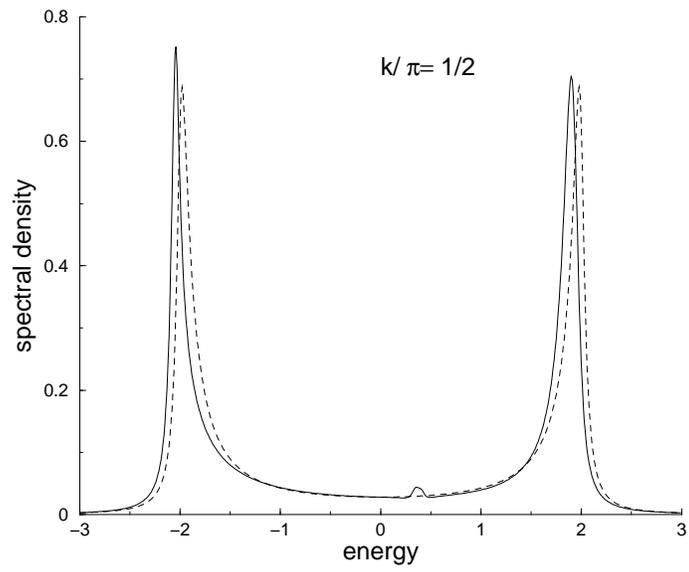
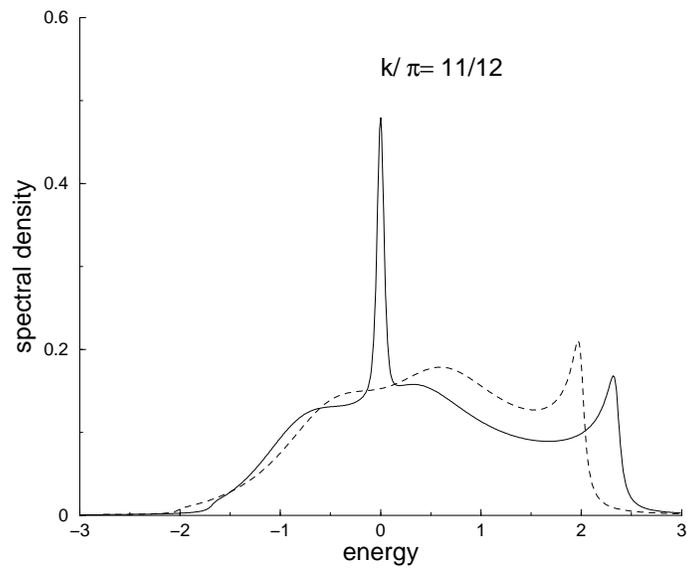
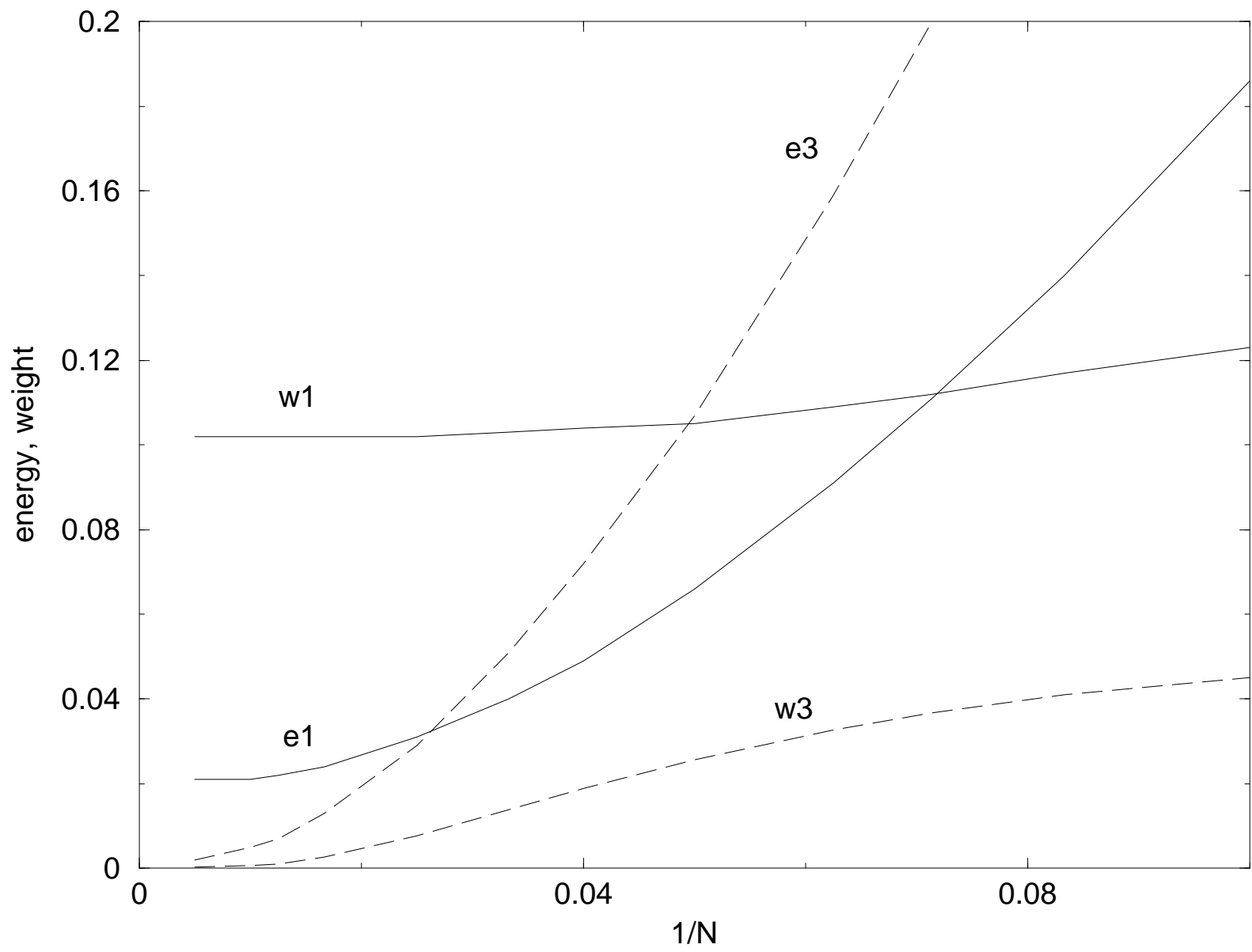
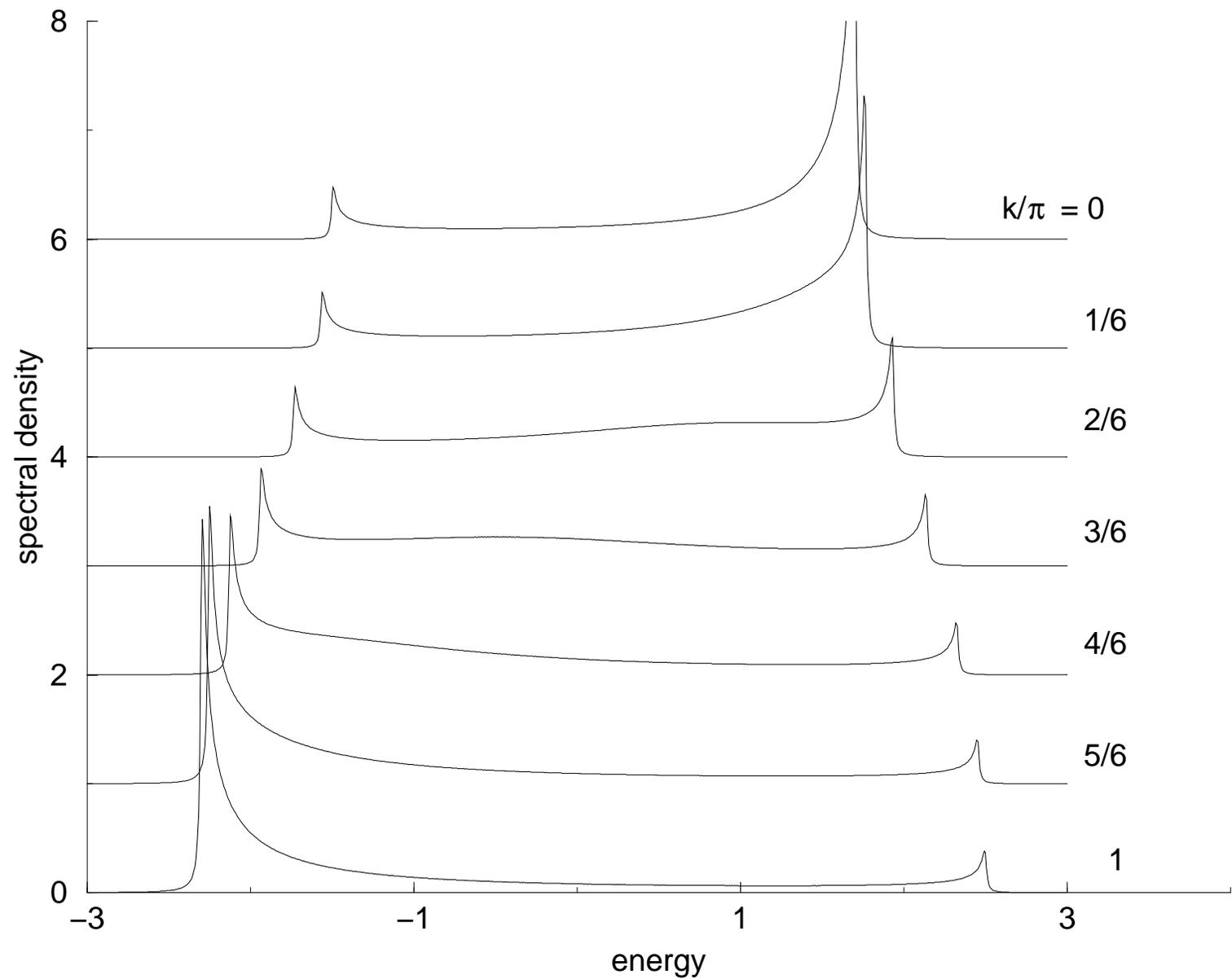
\end{document}